\documentclass[onecolumn,preprintnumbers,amsmath,amssymb]{revtex4}
\usepackage{graphicx}% Include figure files
\usepackage{dcolumn}% Align table columns on decimal point
\usepackage{bm}% bold math

\begin{document}

%\preprint{APS/123-QED}

\title{Looping of a long chain polymer in solution: Simple derivation for exact solution for a delta function sink}
\author{Moumita Ganguly*}
\author{Aniruddha Chakraborty}
\affiliation{School of Basic Sciences, Indian Institute of Technology Mandi,
Kamand Campus, Himachal Pradesh 175005, India.}
\date{\today }
\begin{abstract} 
\noindent A simple analytical method for solving intra-molecular reactions of polymer chain in dilute solution is formulated. The physical problem of looping can be modeled mathematically with the use of a Smoluchowski-like equation with a Dirac delta function sink of finite strength. Here we have proposed a very simple method of solution. The solution is expressed in terms of Laplace Transform of the Green's function for end-to-end motion of the polymer in absence of the sink. We have used two different rate constants, the long term rate constant and the average rate constant. The average rate constant and long term rate constant varies with several parameters such as length of the polymer (N), bond length (b) and the relaxation time ${\tau_R}$. Our results are in agreement with that produced by more general and detailed method.
\end{abstract}
\maketitle

\section{Introduction}
\noindent
The theory of end-to-end polymer looping is one of the most widely acclaimed research area both for  theoretical and experimental chemical scientists. The phenomenon of looping is fundamental and central to many essential biological intra-cellular processes like protein folding  \cite {Hudgins,Buscaqlia}, RNA folding \cite {Thirumalai}. The looping of a polymer molecule in solution has been mathematically represented by a Smoluchowski-like equation with a Dirac delta sink of infinite and finite strength \cite{Szabo,M-Physica,M-CPL}. But the detailed derivation of rate constants in case of sink of finite strength include complicated calculations involving operatpr algebra. In this paper, we derive the same analytical expression for rate constants using a very simple method.

\section{End-to-end motion of one-dimensional polymer}

\noindent Here we use the simplest possible one dimensional description of a polymer as introduced by Szabo {\it et.al.,} \cite{Szabo}. The model consists of a total $2N$ segments of length unity. This chain consists of a total $2N$ monomers doing random walk. As it is one dimensional random walk, so each monomer is allowed to take two possible orientations, one along the right and the other among the left direction. So the polymer can have any of one of $2^{2N}$ different conformations. In the following, we define $x$ as the end-to-end distance, such that $x=2j$. After $N$ number of steps the polymer segments can be either on the right $N+j$ or on the left $N-j$. So it is possible to solve the polymer looping problem using the standard techniques of probability theory. Hence the equilibrium end-to-end distribution $P_{0j}$ of this long polymer chain is given by the following equation.
\begin{equation}
P_{0j}=2^{-2N}(^{2N}_{N+j}).
\end{equation}
Now if we consider a general random walk problem, then the probability distribution can be calculated by knowing the rate of fluctuations between the polymers moving to either side, either right or left. If we imagine the polymer molecule to be immersed in solvent, the motion of polymer will be determined by various intra-molecular and inter-molecular forces between the polymer and the solvent, so the polymer motion would appear to be random. Then assuming the variation of all right and left monomer segments being independent of each other the fluctuation is directed by the following rate equation.
\begin{equation}
\label{2}\frac{d}{dt}\left[^p_n\right]=\frac{1}{\tau_R}\left[
\begin{array}{cc}
-1&1\\
1& -1
\end{array}
\right]\left[^p_n\right],
\end{equation}
where the vector $[^p_n]$ represents the activity of right and left segments orientations. $\tau_R$ is the relaxation time between any two different configurations. Now, entire event of end-to-end looping of a polymer molecule in a solution, can be considered as a very simple random walk model confined in $2^{2N}$-dimensional configuration space. The individual monomer's reorientation would result in $2N$ ways to reorder a $x=2j$ conformation either to a $x=2j+2$ or $x=2j-2$ conformation and $N-J+1$ ways to reorient a $x=2j+2$ conformation into a $x=2j$ conformation. Then the resulting Master equation for the end-to-end distribution $P(j,t)$ in the $(2 N + 1)$-dimensional space is given by \cite{Szabo}
\begin{equation}
\tau_R\frac{d}{dt}P(j,t)=-2NP(j,t)+(N+j+1)P(j+1,t)+(N-j+1)P(j-1,t).
\end{equation}
\noindent As we see that for long chain molecule ($N$ very large), what we try to search for and measure is a distribution, {\it{i.e.,}} the probability for finding the particular end-to-end distance of a long polymer molecule. The equilibrium distribution of  Eq.(1) is generally approximated by the continuous Gaussian distribution ($x = 2 b j$)
\begin{equation}
\label{de}
P_{0}(x)=\frac{e^{-\frac{x^2}{4 b^2 N}}}{(4 \pi b^2 N)^{1/2}}.
\end{equation}
The end-to-end distribution of a long chain polymer molecule in solution can be represented in a continuum limit as in Eq. (\ref{de}) as its equilibrium distribution. Then the corresponding probability conservation equation is given below
\begin{equation}
\tau_{R}\frac{\partial P(x,t)}{\partial t} = 4 N b^2 \frac{\partial^2 P(x,t)}{\partial x^2} + 2 \frac{\partial}{\partial x} x P(x,t).
\end{equation}
In the above `$b$' denotes the bond length of the polymer and $x$ denotes the end-to-end distance.

\section{End-to-end reaction of one-dimensional polymer}

\noindent  Now if the two ends of this long chain polymer molecule in solution come into a particular distance, a loop would form, {\it i.e.,} at $ x = x_c $. The occurrence of the looping reaction is incorporated in our model by adding a $x$ dependent sink term $S(x)$ \cite{M-Physica,M-CPL} (taken to be normalized {\it i.e.,} $\int_{-\infty}^{\infty} S(x)dx = 1 $) in the above equation to get
\begin{equation}
\tau_{R}\frac{\partial P(x,t)}{\partial t} = 4 N b^2 \frac{\partial^2 P(x,t)}{\partial x^2} + 2 \frac{\partial}{\partial x} x P(x,t) - k_{0} S(x) P(x,t).
\end{equation}
The above equation can be used to describe the dynamics of loop formation. The term $P(x,t)$ describes the probability density of end-to-end distance at a given time $t$. In our model, the effect of all other chemical reactions involving at least one of the end group, other than the end-to-end loop formation are incorporated through the $k_{s} P(x,t)$ term.
\begin{equation}
\tau_{R}\frac{\partial P(x,t)}{\partial t} = 4 N b^2 \frac{\partial^2 P(x,t)}{\partial x^2} + 2 \frac{\partial}{\partial x} x P(x,t) - k_{0} S(x) P(x,t) - k_{s} P(x,t).
\end{equation}
The rate $k_{s}$ may be regarded as the rate of loss of probability of end-to-end distance.

\section{Exact analytical result}

\noindent In our earlier paper, we have used a very detailed method involving operator algebra to drive the exact analytical expression for rate constants.
In the following, we give a very simple derivation for the same problem. Now we do the Laplace transform of $P(x,t)$ by using the following formula

\begin{equation}
\tilde P(x,s)= \int^\infty_0 P(x,t) e^{-st} dt.
\end{equation}
Laplace transformation of Eq.(7) gives
\begin{equation}
\left[s {\tilde P}(x,s)-\frac{4 N b^2}{\tau_{R}} \frac{\partial^2{\tilde P}(x,s)}{\partial x^2} - \frac{2}{\tau_{R}}\frac{\partial{\tilde P}(x,s)}{\partial x} x + \frac{k_{0}}{\tau_{R}} S(x){\tilde P}(x,s)+\frac{k_{s}}{\tau_{R}}{\tilde P}(x,s)\right] =  P(x,0).
\end{equation}
In the following we assume that $S(x)$ may be represented as Dirac delta function of arbitrary location {\it i.e., } $S(x) = \delta (x - x_c)$. So Eq. (9) becomes
\begin{equation}
\left[s {\tilde P}(x,s)-\frac{4 N b^2}{\tau_{R}} \frac{\partial^2{\tilde P}(x,s)}{\partial x^2} - \frac{2}{\tau_{R}}\frac{\partial{\tilde P}(x,s)}{\partial x} x + \frac{k_{0}}{\tau_{R}} \delta (x - x_c){\tilde P}(x,s)+\frac{k_{s}}{\tau_{R}}{\tilde P}(x,s)\right] =  P(x,0).
\end{equation}
The Eq.(10) can be further simplified to
\begin{equation}
\left[s {\tilde P}(x,s)-\frac{4 N b^2}{\tau_{R}} \frac{\partial^2{\tilde P}(x,s)}{\partial x^2} - \frac{2}{\tau_{R}}\frac{\partial{\tilde P}(x,s)}{\partial x} x + \frac{k_{0}}{\tau_{R}} \delta (x - x_c){\tilde P}(x_c,s)+\frac{k_{s}}{\tau_{R}}{\tilde P}(x,s)\right] =  P(x,0).
\end{equation}
The solution of this equation using Green's function $G(x,s|x_0)$ is given below
\begin{equation}
\tilde P(x,s)= \int^\infty_{-\infty} dx_{0}G(x,s+\frac{k_{s}}{\tau_{R}}|x_0)P(x_0,0) - \frac{k_{0}}{\tau_{R}} {\tilde P}(x_c,s)\int^{\infty}_{-\infty} dx_{0}G(x,s+\frac{k_{s}}{\tau_{R}}|x_0)\delta(x_0 - x_c).
\end{equation}
After doing integration in the last term of the right hand side of the above equation, we get
\begin{equation}
\tilde P(x,s)= \int^\infty_{-\infty} dx_{0}G(x,s+\frac{k_{s}}{\tau_{R}}|x_0)P(x_0,0) - \frac{k_{0} }{\tau_{R}}{\tilde P}(x_c,s)G(x,s+\frac{k_{s}}{\tau_{R}}|x_c).
\end{equation}
Now we put $x=x_c$ in the above equation to get
\begin{equation}
\tilde P(x_c,s)= \int^\infty_{-\infty} dx_{0}G(x_c,s+\frac{k_{s}}{\tau_{R}}|x_0)P(x_0,0) - \frac{k_{0}}{\tau_{R}}{\tilde P}(x_c,s)G(x_c,s+\frac{k_{s}}{\tau_{R}}|x_c).
\end{equation}
Now we solve the above equation for $\tilde P(x_c,s)$ to get
\begin{equation}
\tilde P(x_c,s)= \frac{\int^\infty_{-\infty} dx_{0}G(x_c,s+\frac{k_{s}}{\tau_{R}}|x_0)P(x_0,0)}{1+\frac{ k_{0}}{\tau_{R}}G(x_c,s+\frac{k_{s}}{\tau_{R}}|x_c)}.
\end{equation}
When we substituted back into Eq. (14) we get
\begin{equation}
\tilde P(x,s)= \int^\infty_{-\infty} dx_{0} \left[G(x,s+\frac{k_{s}}{\tau_{R}}|x_0)- \frac{\frac{ k_0}{\tau_{R}} G(x,s+\frac{k_{s}}{\tau_{R}}|x_c)G(x_c,s+\frac{k_{s}}{\tau_{R}}|x_0)}{1+ \frac{k_0}{\tau_{R}} G(x_c,s+\frac{k_{s}}{\tau_{R}}|x_c)]}\right] P(x_0,0).
\end{equation}
The above equation is exactly the same as reported by us earlier \cite{M-Physica, M-CPL}.
It is difficult to calculate survival probability in time domain $P_e(t) =\int^\infty_{-\infty} dx P(x,t)$. Instead one can easily calculate the Laplace transform of survival probability $P_e(s)$ of $ P_e(t)$ directly. $P_e(s)$ is associated to $P(x,s)$ by 
\begin{equation}
P_e(s) = \int^\infty_{-\infty} dx {\tilde P}(x,s).
\end{equation} 
Using Eq. (17) and the fact that $\int^\infty_{-\infty} dx_{0} (G(x,s|x_0) = 1/s$, we get
\begin{equation}
P_e(s)=\frac{1}{s+k_{s}/\tau_{R}}\left[1-[1+\frac{k_0}{\tau_{R}} G(x_c,s+ k_{s}/\tau_{R}|x_c)]^{-1} \frac{ k_0}{\tau_{R}}\times \int^\infty_{-\infty} dx_0 G (x_c,s+k_{s}/\tau_{R}|x_0)P(x_0,0)\right].
\end{equation}
The average and long time rate constants can be easily evaluated from the formula of $P_e(s).$ Thus, $k^{-1}_I =P_e(0)$ and $k_L$ = negative of the pole of $P_e(s),$ which is close to the origin. From Eq. (19), we obtain
\begin{equation}
k^{-1}_I =\frac{\tau_{R}}{k_{s}}\left[1- [1+\frac{k_0}{\tau_{R}} G(x_c,k_{s}/\tau_{R}|x_c)]^{-1} \frac{k_0}{\tau_{R}} \; \times \int^\infty_{-\infty} dx_0 G(x_c,k_{s}/\tau_{R}|x_0)P(x_0,0)\right].
\end{equation}
Thus $k_I$ depends on the initial probability distribution $P(x,0)$, whereas $k_L = - $ pole of $[\;[ 1+\frac{2 k_0 \epsilon}{\tau_R}\; G(x_c, s+k_s/\tau_{R}|x_c)][s+k_s/\tau_{R}]\;]^{-1}$, the one which is closest to the origin, on the negative $s$ - axis, and is independent of the initial distribution, $P(x_0,0)$. The $G_0(x,s;x_0)$ can be found out by using the following equation \cite{KLS}:
\begin{equation}
\left(s - {\cal L}\right) G_{0}(x,s;x_0)= \delta (x - x_0). 
\end{equation}
Using standard method \cite{Hilbert} to obtain.
\begin{equation}
G_0(x,s;x_0)=F(z,s;z_0)/(s+k_s)
\end{equation}
with
\begin{equation}
F(z,s;z_0)= D_\nu(-z_<)D_\nu(z_>)e^{(z_0^2-z^2)/4}\Gamma(1-\nu)[1/(4 \pi N b^2)]^{1/2}. 
\end{equation}
In the above, $z$ defined by $z = x(2Nb^2)^{1/2}$  and $z_j = x_j(2Nb^2)^{1/2}$, $\nu  = —s{\tau_{R}}/2$ and $\Gamma(\nu)$ is the gamma function. Also, $z_{<}= min(z, z_0)$ and $z_{>}= max(z, z_0)$. $D_{\nu}$ represent parabolic cylinder functions. To understand the behavior of $k_I$ and $k_L$, we assume the initial distribution $P^0_e(x_0)$ is represented by $\delta(x-x_0)$. Then, we get 
\begin{equation}
{k_I}^{-1}= (k_s/\tau_{R})^{-1}\left(1 - \frac{\frac{k_0}{\tau_{R}}F(z_s,k_s/\tau_{R}|z_0)}{k_s/\tau_{R}+ {\frac{k_0}{\tau_{R}}}F(z_s,k_s/\tau_{R}|z_s)} \right).
\end{equation}
Again
\begin{equation}
k_L= \frac{k_s}{\tau_{R}} - [ values \; of \; s \; for \; which \;\; s+ {\frac{k_0}{\tau_{R}}} F(z_s,s|z_s)=0].
\end{equation}
We should mention that $k_I$ is dependent on the initial position $x_0$ and $k_s$ whereas $k_L$ is independent of the initial position. In the following, we consider $\frac{k_s}{\tau_{R}}\rightarrow$ 0, in this limit we get the conclusions, which we expect to be valid even when $k_s$ is finite. Using the properties of $D_v{(z)}$, we find that when $\frac{k_s}{\tau_{R}}\rightarrow 0, F{(z_s,k_s/\tau_{R}|z_0)}$ and $F{(z_s,k_s/\tau_{R}|z_s)}\rightarrow exp(-z_s^2/2){[1/(4\pi Nb^2)]}^\frac{1}{2}$ so that
\begin{equation}
\frac{k_0}{\tau_R} F{(z_s,k_s/\tau{R}|z_0)}/[k_s/\tau_{R}+ \frac{k_0 }{\tau_R}F{(z_s,k_s/\tau_{R}|z_s)}]\rightarrow 1.
\end{equation}
Hence, we get 
\begin{equation}
k_I^{-1}=-{[\frac{\partial}{\partial k_s/\tau_R}\left[\frac{\frac{k_0}{\tau_R} F(z_s,k_s/\tau_R|z_0)}{k_s/\tau_R + \frac{k_0}{\tau_R} F(z_s,k_s/\tau_R|z_s)}\right]}_{(k_s/\tau_R) \rightarrow 0}.
\end{equation} 
If we take $z_0 < z_s $, so that the particle is initially placed to the left side of the sink. Then we get
\begin{equation}
k_I^{-1}= \frac{e^{{z_s}^2/2} \tau_R}{k_0}{[1/{(4\pi N b^2)}]}^{1/2}+ \left[\frac{\partial}{\partial k_s}\left[\frac{e^{[(z_0^2-z_s^2)/4]}D_v{(-z_0)}}{D_v{(-z_s)}}\right]\right]_{v=0}.
\end{equation} 
After simplification we get the following expression
\begin{equation}
k_I^{-1}= \frac{e^{{z_s}^2/2} \tau_R}{k_0}{[1/{(4\pi N b^2)}]}^{1/2}+ \left(\int_{z_0}^{z_s} dz e^{(z^2/2)}\left[1+erf(z/\sqrt{2}\right]\right)(\pi/2)({\tau_R/2}).
\end {equation}
The long-term rate constant $k_L$ is determined by the value of $s$, which satisfy $ s + \frac{ k_0  }{\tau_R}F(z_s,s|z_s)=0 $. This equation can be written as an equation for $\nu (= -s{\tau_R} /2)$
\begin{equation}
\nu = D_\nu(-z_c)D_\nu(z_c)\Gamma(1-\nu)\frac{2 k_0 \epsilon }{4 b \sqrt{\pi N}} 
\end{equation}
For integer values of $\nu$, $D_\nu(z)=2^{-\nu/2}e^{-z^2/4}H_{\nu}(z/\sqrt{2})$, $H_{\nu}$ are Hermite polynomials. $\Gamma(1-\nu)$ has poles at $\nu = 1,2, . . . .$. Our interest is in the case where $\nu \in [0, 1]$, as $k_L = \frac{2}{\tau_R} \nu$ for $k_s =0$. If $ \frac{k_0 }{(4 b \sqrt{\pi N})}\ll 1$, or $z_c \gg 1$ then $\nu \ll 1$ and one can arrive
\begin{equation}
\nu = D_0(-z_s)D_0(z_s)\frac{k_0}{(4 b \sqrt{\pi N)}} 
\end{equation}
and hence 
\begin{equation}
k_L = \frac{k_{0} e^{{-z_s}^2/2}}{\tau_R}{[1/{(4\pi N b^2)}]}^{1/2}.
\end{equation}
In this limit, the rate constant $k_L$ exhibits Arrhenius type activation. Interestingly both the rate constants are directly proportional to the width of the sink.

\section{Conclusions:}

\noindent In this paper, we have proposed a simple analytical method for calculating rate constants of looping of a long polymer molecule in solution. Explicit expressions for $k_{I}$ and $k_{L}$ have been derived. Results obtained using our new method is same as that found earlier by a more detailed and complicated method \cite{M-Physica, M-CPL}.

\section{Acknowledgments:}
\noindent One of the author (M.G.) would like to thank IIT Mandi for HTRA fellowship and the other author (A.C.) thanks IIT Mandi for providing PDA grant.

\end{document}